\documentclass[twocolumn,prb]{revtex4}
\usepackage{graphicx}
\usepackage{amsmath}
\usepackage{amssymb}
\usepackage{bm}

\begin{document}
\newcommand{\etal}{{\it et al. }}
\newcommand{\Tr}{{\rm Tr }}
\newcommand{\Si}{{\rm Si }}
\renewcommand{\vec}[1]{\bm{#1}}
\renewcommand{\Im}{{\rm Im }}
%
\title{Twisted exchange interaction between localized
spins embedded in a one- or two-dimensional electron gas with
Rashba spin-orbit
coupling}

\author{Hiroshi Imamura}
\affiliation{Graduate School of Information Sciences, Tohoku University,
  Sendai 980-8579, Japan}
\author{Patrick Bruno and Yasuhiro Utsumi}
\affiliation{Max-Planck-Institut f{\" u}r Mikrostrukturphysik
Weinberg 2, D-06120 Halle, Germany}
\pacs{}
%

\begin{abstract}
  We study theoretically the Ruderman-Kittel-Kasuya-Yosida (RKKY)
  interaction in one- and two-dimensions in presence of a Rashba
  spin-orbit (SO) coupling.
  We show that rotation of the spin of conduction electrons due to
  SO coupling causes a twisted RKKY interaction between localized spins
  which consists of three
  different terms: Heisenberg, Dzyaloshinsky-Moriya, and Ising
  interactions.
  We also show that the effective spin Hamiltonian reduces to the
  usual RKKY interaction Hamiltonian in the twisted spin space where
  the spin quantization axis of one localized spin is rotated.
\end{abstract}

\maketitle
%
%
There has been a great deal of interest in the field of
spintronics where spin degrees of freedom of electrons are
manipulated to produce a desirable outcome
\cite{maekawa-book,awschalom-book}. Eminent examples are given by
the giant magnetoresistance (GMR) effect
\cite{baibich1988,velu1988,binasch1989} and the interlayer
exchange coupling in magnetic multilayers
\cite{grunberg1986,carbone1987,parkin1990}. The interlayer
exchange coupling is explained in the context of
Ruderman-Kittel-Kasuya-Yosida (RKKY) interaction
\cite{bruno1991,bruno1992}, equivalently, or in terms of
spin-dependent electron confinement \cite{edwards1991,bruno1995}.
The RKKY interaction is an indirect exchange interaction between
two localized spins via the spin polarization of conduction
electrons \cite{ruderman1954,kasuya1956,yosida1957,kittel1968}.

Recently, much attention has been focused on the effect of the
Rashba spin-orbit (SO) coupling in two-dimensional electron gases
(2DEG) \cite{rashba1960}. Investigation of the Rashba effect of
2DEG in semiconductor heterostructures has been stimulated by the
proposition of a spin field effect transistor \cite{datta1990}. It
has been established the Rashba SO coupling can be controlled by
means of a gate voltage
\cite{nitta1997,engels1997,dirk2000,koga2002}. The Rashba effect
has also been observed in 2DEG formed from surface states
electrons at metal surfaces such as Au(111)
\cite{lashell1996,reinert2001,nicolay2002,reinert2003,henk2003},
Li/W(110) or Li/Mo(110) \cite{rotenberg1999}. It has also been
found that confinement of the surface state due to atomic steps on
vicinal surfaces leads to quasi one-dimensional (1D) surface
states, which also exhibit the Rashba effect
\cite{mugarza2001,ortega2002,mugarza2002}.

Usually the RKKY interaction yields a parallel or antiparallel
coupling of localized spins (Heisenberg coupling). However, if
spin of conduction electrons precesses due to the spin-orbit
coupling, it can be possible to produce a non-collinear
Dzyaloshinsky-Moriya (DM) coupling of localized spins
\cite{dzyaloshinsky1958,moriya1960,moriya1960b}. In this paper, we
investigate the RKKY coupling between localized spins embedded in
a 1D- or 2DEG with Rashba SO coupling. We show that rotation of
the spin of conduction electrons due to the Rashba SO coupling
causes a {\em twisted} RKKY interaction between localized spins which
consists of three different terms: Heisenberg,
Dzyaloshinsky-Moriya, and Ising interactions. We point out that a
perturbative treatment of the SO coupling as is usually done
\cite{dzyaloshinsky1958,moriya1960,moriya1960b} is valid only for
small distances between the localized spins; in this case the DM
and Ising terms are respectively linear and quadratic with respect
to the SO coupling strength. In the limit of large distances, a
non-perturbative treatment of the SO coupling is necessary, and
one obtains DM and Ising terms that have the same oscillation
amplitude as the Heisenberg term, independently of the SO coupling
strength. This peculiar behavior of the twisted RKKY interaction
for a pair of localized spins can be explained by introducing a
twisted spin space where the spin quantization axis of one 
of the localized spins is rotated.


%
We consider a system consisting of two localized spins embedded in a
1D- or 2DEG with a Rashba type spin-orbit coupling
\cite{rashba1960}. The Hamiltonian for the conduction electrons is
given by
\begin{equation}
  H_{0}=-\frac{\hbar^{2}}{2m}\vec{\nabla}^{2} +
  \alpha\left(-i\hbar\vec{\nabla}\times\hat{\vec{z}}\right)\cdot\vec{\sigma},
\label{eq:hzero}
\end{equation}
where $\alpha$ represents the strength of the spin-orbit coupling,
$\hat{\vec{z}}$ is a unit vector along the $z$-axis, and
$\vec{\sigma}$ is the vector of Pauli spin matrices.
We assume that the conduction electrons are confined in a wire along
the $x$-axis (one-dimensional system) or in the $x-y$ plane
(two-dimensional system).
The direction of the effective electric field of spin-orbit coupling
is taken to be along the $z$-axis for both one- and two-dimensional
systems.

Since the Hamiltonian $H_{0}$ commutes with the momentum operator
$-i\hbar\vec{\nabla}$, the wave vector $\vec{k}$ is a good
quantum number.  The Green function of the conduction
electrons in the real space can be expressed as
\begin{equation}
G\left(\vec{R};z\right) \equiv \frac{1}{(2\pi)^{D}} \int
d^D\vec{k} \ e^{i \vec{k}\cdot\vec{R}}\ G(\vec{k};z),
\label{eq:gfr}
\end{equation}
where $D=1$ or 2 is the dimension of the system and the Green
function in the momentum space is given by
\begin{equation}
G(\vec{k};z)
=
\left[
  z-
  \left\{
    \frac{\hbar^{2}k^{2}}{2m}\sigma_{0}
    +\alpha\left(\vec{k}\times\hat{\vec{z}}\right)\cdot\vec{\sigma}
  \right\}
\right]^{-1}.
\end{equation}
Here $\sigma_{0}$ is the $(2\times 2)$ unit matrix in the spin
space of conduction electrons.

The localized spins are denoted by $\vec{S}_{1}$ and $\vec{S}_{2}$ and
located at positions $\vec{R}_{1}$ and $\vec{R}_{2}$, respectively.
The coupling between conduction electrons and localized spins is
expressed as the $s$-$d$ interaction Hamiltonian
\begin{equation}
  H_{1}=J\sum_{i=1,2}\delta(\vec{r}-\vec{R}_{i})\vec{S}_{i}\cdot\vec{\sigma},
\end{equation}
where $J$ represents the strength of the $s$-$d$ interaction. Note
that $J$ has the following dimensionality:
(energy)$\times$(length)$^D$.

The total Hamiltonian is given by the sum of $H_{0}$ and $H_{1}$.
We assume that the coupling constant $J$ is so small that we can treat
$H_{1}$ as a perturbation on $H_{0}$.
The RKKY interaction between $\vec{S}_{1}$ and $\vec{S}_{2}$ is
calculated from the second order perturbation theory as
\begin{equation}
  \begin{aligned}
  H_{1,2}^{\rm RKKY}
  &=
  -\frac{1}{\pi}
  \Im
  \int_{-\infty}^{\varepsilon_{F}} d\varepsilon\
  \Tr
  \left[
    \left(\vec{S}_{1}\cdot\vec{\sigma}\right)
    G\left(\vec{R}_{12};\varepsilon+i0^{+}\right)\right.\\
    &\hspace{5em}\left.
    \left(\vec{S}_{2}\cdot\vec{\sigma}\right)
    G\left(-\vec{R}_{12};\varepsilon+i0^{+}\right)
  \right],
  \end{aligned}
  \label{eq:hrkky}
\end{equation}
where $\varepsilon_{F}$ is the Fermi energy,
$\vec{R}_{12}\equiv\vec{R}_{1}-\vec{R}_{2}$, $i0^{+}$ represents
an infinitesimal imaginary energy, and Tr means a trace over the
spin degrees of freedom of conduction
electrons\cite{bruno1995,schwabe1996}.

%
Let us first consider the one-dimensional case. The Green function
of conduction electrons in the momentum space is expressed as
\begin{equation}
 G(k;\varepsilon+i0^{+})
 =
 \left[
   \varepsilon+i0^{+}-
   \left\{
     \frac{\hbar^{2}k^{2}}{2m}\sigma_{0}
     -\alpha k \sigma_{y}
   \right\}
 \right]^{-1}.
 \label{eq:gf1d}
\end{equation}
After some algebras, Eq. (\ref{eq:gf1d}) can be written as
\begin{equation}
 G(k;\varepsilon+i0^{+})
 =
 G_{0}(k;\varepsilon)\sigma_{0} + G_{1}(k;\varepsilon)\sigma_{y},
\label{eq:gfk1d}
\end{equation}
where the diagonal and off-diagonal Green functions are defined as
\begin{align}
  &
  \begin{aligned}
    G_{0}(k;\varepsilon)
    &\equiv
    \frac{m}{\hbar^{2}}
    \left[
      \frac{1}{k_{\varepsilon}^{2} -k^{2} -2k k_{R} + i0^{+}}\right.\\
      &\hspace{4em}\left.
      +
      \frac{1}{k_{\varepsilon}^{2} -k^{2} +2k k_{R} + i0^{+}}
    \right],
  \end{aligned}
\end{align}
\begin{align}
  &\begin{aligned}
    G_{1}(k;\varepsilon)
    &\equiv
    \frac{m}{\hbar^{2}}
    \left[
      \frac{1}{k_{\varepsilon}^{2} -k^{2} -2k k_{R} + i0^{+}}\right.\\
      &\hspace{4em}\left.
      -
      \frac{1}{k_{\varepsilon}^{2} -k^{2} +2k k_{R} + i0^{+}}
    \right].
  \end{aligned}
\end{align}
with $k_{\varepsilon}^{2}\equiv 2 m \varepsilon/\hbar^{2}$ and
$k_{R}\equiv m \alpha / \hbar^{2}$.

A straight-forward contour calculation gives
\begin{equation}
  G(R;\varepsilon+i0^{+})
  =
  G_{0}(R;\varepsilon)\sigma_{0}
  +
  G_{1}(R;\varepsilon)\sigma_{y},
  \label{eq:g1d}
\end{equation}
where
\begin{align}
  G_{0}(R;\varepsilon)
  &=
  -i\frac{ m}{\hbar^{2}(q+i0^{+})}e^{i q |R|}
  \cos\left(k_{R} R\right) ,
  \label{eq:g1d0}\\
  G_{1}(R;\varepsilon)
  &=
  \frac{m}{\hbar^{2}(q +i0^{+})}e^{i q |R|}
  \sin\left(k_{R} R\right),
  \label{eq:g1d1}
\end{align}
with
\begin{equation}
q\equiv\sqrt{k_{\varepsilon}^{2}+k_{R}^{2}}=\sqrt{\frac{2m}{\hbar^{2}}\varepsilon
+ k_{R}^{2}}.
\label{eq:qdef}
\end{equation}

Substituting Eqs. (\ref{eq:g1d})-(\ref{eq:g1d1}) into Eq.
(\ref{eq:hrkky}), and using the relations
$G_{0}(-R;\varepsilon)=G_{0}(R;\varepsilon)$ and
$G_{1}(-R;\varepsilon)=-G_{1}(R;\varepsilon)$, we have

\begin{widetext}
\begin{equation}
\begin{aligned}
H_{1,2}^{\rm RKKY}
&=
-\frac{1}{\pi}J^{2}
\Im
\Biggl[
\Tr
\left\{
  \left(\vec{S}_{1}\cdot\vec{\sigma}\right)
  \left(\vec{S}_{2}\cdot\vec{\sigma}\right)
\right\} \int_{-\infty}^{\varepsilon_{F}} d\varepsilon\
G_{0}(R_{12};\varepsilon)^{2} + \Tr \left\{
  \left(\vec{S}_{1}\cdot\vec{\sigma}\right)\sigma_{y}
  \left(\vec{S}_{2}\cdot\vec{\sigma}\right)
\right\}
\int_{-\infty}^{\varepsilon_{F}} d\varepsilon\
G_{1}(R_{12};\varepsilon)G_{0}(R_{12};\varepsilon)\\
&
-\Tr
\left\{
  \left(\vec{S}_{1}\cdot\vec{\sigma}\right)
  \left(\vec{S}_{2}\cdot\vec{\sigma}\right)\sigma_{y}
\right\} \int_{-\infty}^{\varepsilon_{F}} d\varepsilon\
G_{0}(R_{12};\varepsilon)G_{1}(R_{12};\varepsilon) -\Tr \left\{
  \left(\vec{S}_{1}\cdot\vec{\sigma}\right)\sigma_{y}
  \left(\vec{S}_{2}\cdot\vec{\sigma}\right)\sigma_{y}
\right\}
  \int_{-\infty}^{\varepsilon_{F}} d\varepsilon\  G_{1}(R_{12};\varepsilon)^{2}
\Biggr],
\end{aligned}
\end{equation}
\end{widetext}
The traces over the spin operators can be calculated by using the
relation $(\vec{A}\cdot\vec{\sigma})(\vec{B}\cdot\vec{\sigma}) =
(\vec{A}\cdot\vec{B})\sigma_{0} +
i(\vec{A}\times\vec{B})\cdot\vec{\sigma}$, repeatedly
\cite{sakurai-book}:
\begin{align}
  &\Tr
  \left\{
    \left(\vec{S}_{1}\cdot\vec{\sigma}\right)
    \left(\vec{S}_{2}\cdot\vec{\sigma}\right)
  \right\}
  =
  2\vec{S}_{1}\cdot\vec{S}_{2},\\
  &\Tr
  \left\{
    \left(\vec{S}_{1}\cdot\vec{\sigma}\right)\sigma_{y}
    \left(\vec{S}_{2}\cdot\vec{\sigma}\right)
  \right\}
  =-2i(\vec{S}_{1}\times\vec{S}_{2})_{y},\\
  &\Tr
  \left\{
    \left(\vec{S}_{1}\cdot\vec{\sigma}\right)
    \left(\vec{S}_{2}\cdot\vec{\sigma}\right)\sigma_{y}
  \right\}
  =2i(\vec{S}_{1}\times\vec{S}_{2})_{y},\\
  &\Tr
  \left\{
    \left(\vec{S}_{1}\cdot\vec{\sigma}\right)\sigma_{y}
    \left(\vec{S}_{2}\cdot\vec{\sigma}\right)\sigma_{y}
  \right\}
  =2(2S_{1}^{y}S_{2}^{y} - \vec{S}_{1}\cdot\vec{S}_{2}).
  \label{eq:trace}
\end{align}
Thus, using Eqs. (\ref{eq:g1d0})-(\ref{eq:trace}) we find
\begin{equation}
  \begin{aligned}
  &
  H_{1,2}^{\rm RKKY}=
  F_{1}(|R_{12}|)
  \Biggl[
  \cos(2k_{R} R_{12})\vec{S}_{1}\!\cdot\!\vec{S}_{2}\\
  &
  +
  \sin(2k_{R} R_{12})(\vec{S}_{1}\!\times\!\vec{S}_{2})_{y}
  +
  \left\{1-\cos(2k_{R} R_{12})\right\}\vec{S}_{1}^{y}\vec{S}_{2}^{y}
  \Biggr],
  \end{aligned}
  \label{eq:rkky1d}
\end{equation}
where the range function $F_{1}(|R|)$ is defined as
\begin{equation}
  F_{1}(|R|)
  \equiv
  \frac{2J^{2}}{\pi} \left(\frac{m}{\hbar^{2}}\right)^{2}\
  \Im\ \int_{-\infty}^{\varepsilon_{F}}
  \frac{e^{2iq|R|}}{\left(q+i0^{+}\right)^{2}}\
  d\varepsilon.
  \label{eq:range}
\end{equation}
Performing the change of variable $\varepsilon \to q$ and using
standard complex-plane integration techniques, one eventually
obtains:
\begin{equation}
    F_{1}(|R|)=
  \frac{2 J^{2}}{\pi}
  \frac{m}{\hbar^{2}}
  \left[\Si(2q_{F} |R|) - \frac{\pi}{2}\right],
  \label{eq:rkky1df}
\end{equation}
where $q_{F}\equiv\sqrt{2m\varepsilon_{F} / \hbar^{2} + k_{R}^{2}}$
and $\Si(\ )$ is the sine integral function
\cite{gradshteyn-ryzhik}. The range function of  Eq.
(\ref{eq:rkky1df}) is the same form as that of the usual
one-dimensional RKKY interaction\cite{yafet1987,litvinov1998}
except that the Fermi wave vector 
$k_{F}(\equiv\sqrt{2m\varepsilon_{F} / \hbar^{2}})$ is
replaced by $q_{F}$ .

As shown in Eq. (\ref{eq:rkky1d}), the resulting RKKY interaction
consists of three physically quite different interactions:
Heisenberg, Dzyaloshinsky-Moriya and, Ising interactions.  The
Heisenberg and Ising couplings favor a collinear alignment of
localized spins.  On the contrary, the DM coupling favors a
non-collinear alignment of localized spins. For distances (more
precisely, for $k_R|R_{12}|\ll 1$), the DM and Ising terms are,
respectively, linear and quadratic in the Rashba SO coupling
$\alpha$; this corresponds to the result obtained from a
perturbative treatment of the SO coupling. However, for large
distances ($k_R|R_{12}|\gg 1$), a perturbative treatment of the SO
coupling would completely fail: indeed, in this regime, one finds
that the DM and Ising couplings oscillate with the same amplitude
as the Heisenberg term.

This peculiar twisted coupling of localized spins can be easily
understood by introducing the twisted spin space where the spin
quantization axis of the second localized spin $\vec{S}_{2}$ is
rotated by an angle $\theta_{12}=2k_R R_{12}$ around the $y$-axis.
The spin operators for the second localized spin in the twisted
spin space are given by
\begin{align}
  S_{2}^{x}(\theta_{12})
  &=
  \cos\theta_{12}\ S_{2}^{x} + \sin\theta_{12}\ S_{2}^{z},
  \label{eq:sx}
  \\
  S_{2}^{y}(\theta_{12})
  &=S_{2}^{y},\\
  S_{2}^{z}(\theta_{12})
  &=
  \cos\theta_{12}\ S_{2}^{z} - \sin\theta_{12}\ S_{2}^{x}.
  \label{eq:sz}
\end{align}
From the above equations, one can easily show that the inner
product of $\vec{S}_{1}$ and $\vec{S}_{2}(\theta_{12})$ is
\begin{equation}
  \begin{aligned}
  \vec{S}_{1}\cdot\vec{S}_{2}(\theta_{12})
  &=  \cos\theta_{12}\vec{S}_{1}\!\cdot\!\vec{S}_{2}
  +
  \sin\theta_{12}(\vec{S}_{1}\!\times\!\vec{S}_{2})_{y}\\
  &\hspace{1em}+
  \left\{1-\cos\theta_{12}\right\}\vec{S}_{1}^{y}\vec{S}_{2}^{y},
  \end{aligned}
\end{equation}
so that the RKKY interaction of Eq. (\ref{eq:rkky1d}) can be
expressed as
\begin{equation}
  H_{1,2}^{\rm RKKY}
  =
  F_1(|R_{12}|)\vec{S}_{1}\cdot\vec{S}_{2}(\theta_{12}).
  \label{eq:twist}
\end{equation}
Eq.(\ref{eq:twist}) shows that in presence of spin-orbit coupling
the RKKY interaction results in a collinear coupling of localized
spins in the $\theta_{12}$-twisted spin space.

%
%
Next we address the two-dimensional case. The Green function of
conduction electrons now takes the form
\begin{equation}
  G(\vec{R};\varepsilon+i0^{+})
  =
  G_{0}(R;\varepsilon)\sigma_{0}
  +
  G_{1}(R;\varepsilon)(\hat{\vec{z}}\times\hat{\vec{R}})\cdot\vec{\sigma},
  \label{eq:g2d}
\end{equation}
where $R\equiv \|\vec{R}\|$ and $\hat{\vec{R}}\equiv {\vec{R}}/R$
is the unit vector parallel to $\vec{R}$. The Green functions
$G_{0}(R;\varepsilon)$ and $G_{1}(R;\varepsilon)$ can be
calculated in the similar way as in Ref. \cite{dugaev1994}. A
straight forward calculation yields
\begin{align}
  &
  \begin{aligned}
    G_{0}(R;\varepsilon)
    &\!=\!
    -\frac{i m}{4\hbar^{2}}\!
    \left[\!
      \left(\!1\!+\!\frac{k_{R}}{q}\!\right)
      \!H_{0}^{(1)}\!
      \left[\left(q+k_{R}+i0^{+}\!\right)\!R\right]\right.\\
      &\hspace{2em}\left.
      +
      \left(\!1\!-\!\frac{k_{R}}{q}\!\right)
      \!H_{0}^{(1)}\!
      \left[\left(q-k_{R}+i0^{+}\!\right)\!R\right]
     \!\right],
  \end{aligned}
  \\
  &\begin{aligned}
    G_{1}(R;\varepsilon)
    &\!=\!
    -\frac{m}{4\hbar^{2}}\!
    \left[\!
      \left(\!1\!+\!\frac{k_{R}}{q}\!\right)
      \!H_{1}^{(1)}\!
      \left[\left(q+k_{R}+i0^{+}\!\right)\!R\right]\right.\\
    &\hspace{2em}\left.
      -
      \left(\!1\!-\!\frac{k_{R}}{q}\!\right)
      \!H_{1}^{(1)}\!
      \left[\left(q-k_{R}+i0^{+}\!\right)\!R\right]
    \!\right],
  \end{aligned}
\end{align}
where $H_{0}^{(1)}[\ ]$ and $H_{0}^{(1)}[\ ]$ are Hankel functions
\cite{gradshteyn-ryzhik}. Hereafter, we restrict ourselves to the
region $q R \gg 1$ and $k_{R} \ll q$. In this case we can use the
asymptotic form of Hankel's functions \cite{gradshteyn-ryzhik},
\begin{equation}
  H_{n}^{(1)}(z)\simeq\sqrt{\frac{2}{\pi z}}\
  e^{i\left(z-\frac{n\pi}{2}-\frac{\pi}{4}\right)}
  \ \ \ \left(|z|\to \infty\right).
\end{equation}
Thus we have
\begin{align}
  G_{0}(R;\varepsilon)
  &\simeq
  -i\frac{m}{\hbar^{2}}\frac{1}{\sqrt{2\pi q R}}
  e^{i\left(q R -\frac{\pi}{4}\right)} \cos\left(k_{R}R\right),
  \label{eq:gr2d1}
  \\
  G_{1}(R;\varepsilon)
  &\simeq
  \frac{m}{\hbar^{2}}\frac{1}{\sqrt{2\pi q R}}
  e^{i\left(q R -\frac{\pi}{4}\right)} \sin\left(k_{R}R\right).
  \label{eq:gr2d2}
\end{align}

Without restriction, we can take the coordinate system so that the
vector $\vec{R}$ is aligned with the $x$-axis, i. e., $\vec{R}=R\
\hat{\vec{x}}$.  Then the Green function takes the form
\begin{equation}
  G(\vec{R};\varepsilon+i0^{+})
  =
  G_{0}(R;\varepsilon)\sigma_{0}
  +
  G_{1}(R;\varepsilon)\sigma_{y}.
\end{equation}
The RKKY interaction can be obtained in the similar way as
one-dimensional system, and one gets (for $q_F R \gg 1$)
\begin{equation}
  H_{1,2}^{\rm RKKY}
  \simeq
  F_2(R_{12})
  \vec{S}_{1}\cdot\vec{S}_{2}(\theta_{12}) ,
  \label{eq:rkky2df}
\end{equation}
where the range function $F_2(R)$ is given by
\begin{equation}
  F_2(R)
  \simeq
  -\frac{J^{2}}{2\pi^{2}}
  \frac{m}{\hbar^{2}}\frac{\sin\left(2 q_{F} R\right)}{R^{2}}.
  \label{eq:range2}
\end{equation}
Eq. (\ref{eq:rkky2df}) is the same as the usual two-dimensional
RKKY interaction \cite{fischer1975,litvinov1998} except that
$k_{F}$ and $\vec{S}_{2}$ are replaced by $q_{F}$ and
$\vec{S}_{2}(\theta_{12})$, respectively. It is reasonable that
the twisted coupling of two localized spins takes the same form
$\vec{S}_{1}\cdot\vec{S}_{2}(\theta_{12})$ as for the 1D system,
because  for $q_F R \gg 1$ a scattering wave of 2D system behaves
like a plane wave.

%
The 2DEG from surface states at metallic surfaces are good
candidates for investigating the twisted RKKY interaction.
Experiments on Au(111) surfaces have yielded $q_F\simeq
0.17$~\AA$^{-1}$ and $k_R\simeq 0.012$~\AA$^{-1}\ $\cite{reinert2003}; in complete analogy to the effect of negative
gate voltage for 2DEGs at a semiconductor heterojunction
\cite{nitta1997,engels1997}, the adsorption of a noble gas (e.g.,
Xe) produces an effective repulsive potential (because of the
Pauli exclusion principle) and leads to a decrease of the average
Fermi wavevector ($q_F\simeq 0.155$~\AA$^{-1}$) and an increase of
the Rashba splitting ($k_R\simeq 0.015$~\AA$^{-1}$)
\cite{reinert2003}. Furthermore, quasi-1D surface states can also
be obtained for vicinal surfaces, such as Au(788) and
Au(23~23~21), with comparable Rashba splitting
\cite{mugarza2001,ortega2002,mugarza2002}. The distance
between magnetic adatoms deposited on such surfaces can be
controlled either directly by atom manipulation using the tip of a
scanning tunneling microscope, or by exploiting self-organization
processes. Such systems therefore constitute a versatile
laboratory to investigate surface states mediated RKKY
interactions under the influence of the Rashba effect. With $k_R
\simeq 0.015$~\AA$^{-1}$ and a distance $R=10$~\AA\ the twist
angle $\theta=2k_R R$ is of the order of $17^{\rm o}$, which is quite
sizeable. In particular, due to the twisted nature of the RKKY
interaction, very interesting frustration phenomena may be
anticipated.

In semiconductor heterostructures the twisted RKKY interaction may
also be of great interest, in particular in view of possibilities
for manipulating entanglement between spins of quantum dots connected by a
wire with Rashba SO coupling \cite{hima2003}, as needed for a
spin-based solid-state quantum computer
\cite{loss1998,burkard1999}. Considering the value of the Rashba
coupling reported for
In$_{0.53}$Ga$_{0.47}$As/In$_{0.52}$Al$_{0.48}$As heterostructure
{\cite{nitta1997} one obtains that the twist angle $\theta$ can be
controlled from $\theta=\pi$ to $\theta=3\pi/2$ by a gate voltage
for $R=400$nm.

%
In conclusion, we have studied the twisted RKKY interaction in
one- and two-dimensions in presence of Rashba spin-orbit coupling.
We have also shown that in the twisted spin space where the spin
quantization axis of one localized spin is rotated, the twisted
RKKY interaction is expressed in the same form as the usual RKKY
interaction.  The angle $\theta$ between the localized spins can
be controlled by the distance between localized spins $R$ and/or
the strength of the spin-orbit coupling of conduction electrons.

%
We would like to thank Jan Martinek for valuable discussions. This
work was supported by MEXT, Grant-in-Aid for Scientific Research
on the Priority Area "Semiconductor Nanospintronics" No. 14076204,
Grant-in-Aid for Scientific Research (C), No. 14540321 and NAREGI
Nanoscience Project, as well as by the BMBF (Grant No.~01BM924).

%


\end{document}